# Semiparametric regression and risk prediction with competing risks data under missing cause of failure

Giorgos Bakoyannis[1] · Ying Zhang[2] · Constantin T. Yiannoutsos[1]

## Abstract

The cause of failure in cohort studies that involve competing risks is frequently incompletely observed. To address this, several methods have been proposed for the semiparametric proportional cause-specific hazards model under a missing at random assumption. However, these proposals provide inference for the regression coefficients only, and do not consider the infinite dimensional parameters, such as the covariate-specific cumulative incidence function. Nevertheless, the latter quantity is essential for risk prediction in modern medicine. In this paper we propose a unified framework for inference about both the regression coefficients of the proportional cause-specific hazards model and the covariate-specific cumulative incidence functions under missing at random cause of failure. Our approach is based on a novel computationally efficient maximum pseudo-partial-likelihood estimation method for the semiparametric proportional cause-specific hazards model. Using modern empirical process theory we derive the asymptotic properties of the proposed estimators for the regression coefficients and the covariate-specific cumulative incidence functions, and provide methodology for constructing simultaneous confidence bands for the latter. Simulation studies show that our estimators perform well even in the presence of a large fraction of missing cause of failures, and that the regression coefficient estimator can be substantially more efficient compared to the previously proposed augmented inverse probability weighting estimator. The method is applied using data from an HIV cohort study and a bladder cancer clinical trial.

**Keywords** Cause-specific hazard · Cumulative incidence function · Confidence band

✉ Giorgos Bakoyannis
  gbakogia@iu.edu

[1] Department of Biostatistics, Indiana University Fairbanks School of Public Health and School of Medicine, 410 West 10th Street, Suite 3000, Indianapolis, IN 46202, USA

[2] Department of Biostatistics, University of Nebraska Medical Center, Omaha, USA

**Mathematics Subject Classification** 62N01 · 62N02

## 1 Introduction

There is an increasing frequency of epidemiological studies and clinical trials that involve a large number of subjects, longer observation periods and multiple outcomes or competing risks (Ness et al. 2009). The basic identifiable quantities from studies with competing risks are the cause-specific hazard and the cumulative incidence function (Putter et al. 2007; Bakoyannis and Touloumi 2012). Choosing the most relevant estimand in a given study depends on the scientific question of interest: if the goal of the study is to identify risk factors of the competing risks under consideration, the cause-specific hazard is the most relevant quantity (Koller et al. 2012); if the interest is focused on clinical prediction or prognosis, as for example in studies of quality of life, the cumulative incidence function is the most relevant estimand (Fine and Gray 1999; Koller et al. 2012; Andersen et al. 2012).

A frequent problem in studies with competing risks is that cause of failure is incompletely observed, and several methods have been proposed to address this issue under a missing at random assumption. Craiu and Duchesne (2004) proposed an EM-algorithm for estimation under a piecewise-constant hazards competing risks model, for situations with masked cause of failure. Goetghebeur and Ryan (1995) proposed a partial likelihood-based approach for estimating the regression coefficients of the semiparametric proportional cause-specific hazards model under missing cause of failure, by assuming that the baseline hazards for the different causes of failure are proportional. Lu and Tsiatis (2001) proposed a multiple-imputation approach based on a parametric assumption regarding the probability of the cause of failure conditional on the fully observed data. Lu and Tsiatis approach, unlike the estimator by Goetghebeur and Ryan (1995), did not impose the proportionality assumption between the baseline hazards for the different causes of failure. Gao and Tsiatis (2005) developed augmented inverse probability weighting estimators (AIPW) for the regression coefficients in the class of semiparametric linear transformation models. This approach utilizes parametric models for the probability of missingness and the probability of the cause of failure conditional on the fully observed data. Hyun et al. (2012) applied the AIPW approach to the proportional cause-specific hazards model. These AIPW estimators are more efficient compared to the simple inverse probability weighting estimators, and possess the double-robustness property. The latter property ensures consistency even if one of the parametric models for the probability of missingess and the cause of failure probability is incorrectly specified. Recently, Nevo et al. (2018) proposed an estimation approach for the proportional cause-specific hazards model that utilized auxiliary covariates for a weaker missing at random assumption. However, this approach considered an unspecified baseline hazard for only one cause of failure, say $\lambda_{0,1}(t)$, while the baseline hazards for the remaining cause of failures were assumed to satisfy a parametric hazard ratio $\lambda_{0,j}(t)/\lambda_{0,1}(t)$. On the contrary, the other approaches mentioned above considered unspecified baseline cause-specific hazards for all the cause of failures (Lu and Tsiatis 2001; Gao and Tsiatis 2005; Hyun et al. 2012). It is important to note that, none of the aforementioned methods have considered the problem of inference for the infinite-dimensional parameters, such as the covariate-specific cumulative incidence function. However, these personalized risk predictions provide crucial information to clinicians and policy makers in medical decision making and implementation science, as in our motivating study described below.

Several other approaches have been proposed for the semiparametric additive cause-specific hazards model with missing causes of failure (Lu and Liang 2008; Bordes et al. 2014). In this article we focus on the semiparametric proportional cause-specific hazards model because this is the standard model for estimating risk factor effects in practice (Koller et al. 2012). Additionally, other approaches have been proposed for semiparametric models of the cumulative incidence function (Bakoyannis et al. 2010; Moreno-Betancur and Latouche 2013) with missing cause of failure. However, it is more appropriate to analyze the cause-specific hazard function for evaluating risk factors, than analyzing the cumulative incidence function (Koller et al. 2012).

An important gap in the literature of competing risks data with missing cause of failure is that there is currently no unified approach available for inference about both the cause-specific hazard, for evaluating risks factors, and the covariate-specific cumulative incidence function, for risk prediction purposes. Such an approach would be very useful to an ongoing study with competing risks from the East Africa Regional Consortium of the International Epidemiology Databases to Evaluate AIDS (EA-IeDEA). Among other data, EA-IeDEA records death and disengagement from care, the two major outcomes experienced by HIV-infected individuals who receive antiretroviral treatment (ART). The goal of the motivating study is twofold: (1) to identify risk factors of disengagement from HIV care and death in patients who receive ART, and (2) to provide individualized (i.e., covariate-specific) prognosis and prediction estimates for the aforementioned competing risks. The first goal aims at providing a scientific understanding of the factors that are related to disengagement from care and death under ART, while the second goal focuses on informing clinical practice and implementation science efforts to optimize care in a cost-efficient way (Hirschhorn et al. 2007). Therefore, the first goal is focused on making inference about the regression coefficients in a model for the cause-specific hazard functions (Koller et al. 2012), while for the second goal the focus is in covariate-specific cumulative incidence functions (Koller et al. 2012; Andersen et al. 2012). A major complication in the EA-IeDEA study is the significant under-reporting of death. This means that a patient who has been lost to clinic (failure from any cause in our example), could be either dead (whose death has not been reported) or has disengaged from HIV care. Ascertainment of the cause of failure in this study requires intensive outreach of the patients who have been identified as lost to clinic in the community, and subsequent ascertainment of their vital status. However, this is a difficult and costly process and, thus, it is only carried out for a small subset of patients who have been flagged as lost to clinic. This leads to a significant missing cause of failure problem.

In this work, we propose a unified framework for inference about both the regression coefficients and the covariate-specific cumulative incidence functions under the semiparametric proportional cause-specific hazards model with incompletely observed cause of failure. To the best of our knowledge, inference about the covariate-specific cumulative incidence function has not been studied in the literature of missing cause of failure under the semiparametric proportional cause-specific hazards model and the class of linear transformation models. In this article we fill this significant gap in the literature. Our approach is based on a novel computationally efficient maximum pseudo-partial-likelihood estimation (MPPLE) method under the common missing at random assumption. Our estimator utilizes a parametric model for the probability of the cause of failure, which includes auxiliary covariates in order to make the missing at random assumption more plausible (Lu and Tsiatis 2001; Nevo et al. 2018; Bakoyannis et al. 2019). The parametric assumption for the latter model is evaluated through a formal goodness of fit procedure based on a cumulative residual process, similarly to the work by Bakoyannis et al. (2019). Computation of the proposed MPPLE is easily implemented using the function coxph of the R package survival as illustrated in the Electronic Supplementary Material. However, computation of standard errors requires bootstrap methods as we have not implemented the standard error estimators for general use in the R software yet. Using modern empirical process theory, we establish the asymptotic properties of our estimators for both the regression coefficients and the covariate-specific cumulative incidence functions, and propose closed-form variance estimators based on the empirical versions of the corresponding influence functions. In addition, we also propose a method to construct simultaneous confidence bands for the covariate-specific cumulative incidence functions. The finite sample properties of the estimators and their robustness against misspecification of the parametric model for the probability of the cause of failure are investigated through simulations. Moreover, in the simulation studies, we also demonstrate superior finite sample performance of our estimator for the regression coefficients compared to the AIPW estimator (Gao and Tsiatis 2005; Hyun et al. 2012). Finally, we apply the methodology to data sets from the EA-IeDEA HIV cohort study and a bladder cancer trial from the European Organisation for Research and Treatment of Cancer (EORTC).

The rest of the paper is organized as follows: Section 2 provides notation and assumptions that pertain to the model associated with the observed data. Section 3 describes the proposed estimator and its large sample properties. We conduct a number of simulation studies in Sect. 4 by which we justify numerically the validity of the proposed method and compare it with the AIPW method in terms of their finite sample performance. In Sect. 5 the method is applied to the HIV/AIDS study and the bladder cancer trial. We summarize the results and discuss potential extensions of the proposed methodology in Sect. 6. R code, asymptotic theory proofs, and simulation results regarding the infinite-dimensional parameters are provided in the Electronic Supplementary Material.

## 2 Notation and assumptions

Let $T$ and $U$ denote the failure and right censoring times. The corresponding observable quantities are $X = T \wedge U$ and $\Delta = I(T \leq U)$. Additionally, let $C \in \{1, \ldots, k\}$ denote the cause of failure, where $k$ is finite. We assume that the observation interval is $[0, \tau]$, with $\tau < \infty$. Let $\mathbf{Z}$ denote a $p$-dimensional vector of covariates. As mentioned in the Introduction, the basic identifiable quantities from competing risks data are the

cause-specific hazards

$$\lambda_j(t; \mathbf{z}) = \lim_{h \downarrow 0} \frac{1}{h} P(t \leq T < t+h, C = j | T \geq t, \mathbf{Z} = \mathbf{z}), \quad j = 1, \ldots, k$$

and the cumulative incidence functions

$$F_j(t; \mathbf{z}) = P(T \leq t, C = j | \mathbf{Z} = \mathbf{z})$$
$$= \int_0^t \exp\left[-\sum_{l=1}^k \Lambda_l(s; \mathbf{z})\right] \lambda_j(s; \mathbf{z}) ds, \quad j = 1, \ldots, k, \quad (1)$$

where $\Lambda_j(t; \mathbf{z}) = \int_0^t \lambda_j(s; \mathbf{z}) ds$, which is the covariate-specific cumulative hazard for the $j$th cause of failure. A standard model for the cause-specific hazard is the proportional hazards model

$$\lambda_j(t; \mathbf{Z}) = \lambda_{0,j}(t) \exp(\boldsymbol{\beta}_{0,j}^T \mathbf{Z}), \quad j = 1, \ldots, k, \quad (2)$$

where $\lambda_{0,j}(t)$ is the $j$th unspecified baseline cause-specific hazards function for $j = 1, \ldots, k$. Note that, unlike in Nevo et al. (2018), we do not impose further assumptions on the baseline hazards. For competing risks data with incompletely observed cause of failure, we define a missingness indicator $R$, with $R = 1$ indicating that the cause of failure has been observed, and $R = 0$ otherwise. Along with $\mathbf{Z}$, we can potentially observe a vector of auxiliary covariates $\mathbf{A} \in \mathbb{R}^q$, which are not of scientific interest, but may be related to the probability of missingness. Accounting for such auxiliary covariates can make the missing at random assumption more plausible in practice (Lu and Tsiatis 2001; Nevo et al. 2018; Bakoyannis et al. 2019). Throughout this paper, we assume that the event indicator $\Delta$ is always observed and if $\Delta = 0$, we set $R = 1$. Therefore, the observable data $\mathbf{D}_i$ with missing cause of failure are $n$ independent copies of $(X_i, \Delta_i, \Delta_i R_i C_i, \mathbf{Z}_i, \mathbf{A}_i, R_i)$, where $C_i$ is observable only when $\Delta_i = 1$ and $R_i = 1$. Based on the observable data we can define the counting process and at-risk process as $N_i(t) = I(X_i \leq t, \Delta_i = 1)$ and $Y_i(t) = I(X_i \geq t)$ respectively. Additionally, we define the cause-specific counting process as $N_{ij}(t) = I(X_i \leq t, \Delta_{ij} = 1) = \Delta_{ij} N_i(t)$, where $\Delta_{ij} = I(C_i = j, \Delta_i = 1)$ for $j = 1, \ldots, k$, which can only be observed if $R_i = 1$.

In this work, we impose the missing at random assumption $P(R_i = 1 | C_i, \Delta_i = 1, \mathbf{W}_i) = P(R_i = 1 | \Delta_i = 1, \mathbf{W}_i)$, where $\mathbf{W}_i = (T_i, \mathbf{Z}_i, \mathbf{A}_i)$. Note that $T_i$ is observable if $\Delta_i = 1$ since, in this case, $X_i = T_i$. This assumption is equivalent to

$$P(C_i = j | R_i = 1, \Delta_i = 1, \mathbf{W}_i) = P(C_i = j | R_i = 0, \Delta_i = 1, \mathbf{W}_i)$$
$$= P(C_i = j | \Delta_i = 1, \mathbf{W}_i)$$
$$\equiv \pi_j(\mathbf{W}_i, \boldsymbol{\gamma}_0), \quad j = 1, \ldots, k.$$

As in previous work on missing cause of failure in the competing risks model, we assume a parametric model $\pi_j(\mathbf{W}_i, \boldsymbol{\gamma}_0)$ for the $j$th cause of failure, where $\boldsymbol{\gamma}_0$ is a finite-dimensional parameter. A natural choice for $\pi_j(\mathbf{W}_i, \boldsymbol{\gamma}_0)$, $j = 1, \ldots, k$, is the multinomial logit model with the generalized logit link function, if $k > 2$, or the binary logit model with the logit link function, if $k = 2$. In this article, the inverse of the link function for the generalized linear model assumed for $\pi_j(\mathbf{W}_i, \boldsymbol{\gamma}_0)$ is denoted by $g$. For the special case of the binary logit model (whose link function is the logit link), $g$ is the expit function, that is

$$\pi_1(\mathbf{W}_i, \boldsymbol{\gamma}_0) = g[\boldsymbol{\gamma}_0^T(1, \mathbf{W}_i^T)^T] = \frac{\exp[\boldsymbol{\gamma}_0^T(1, \mathbf{W}_i^T)^T]}{1 + \exp[\boldsymbol{\gamma}_0^T(1, \mathbf{W}_i^T)^T]},$$

where $(1, \mathbf{W}_i^T)^T$ is the covariate vector for the $i$th individual that also includes a unit for the intercept, where $\pi_2(\mathbf{W}_i, \boldsymbol{\gamma}_0) = 1 - \pi_1(\mathbf{W}_i, \boldsymbol{\gamma}_0)$.

In this paper, as in Lu and Tsiatis (2001), we assume that the parametric model $\pi_j(\mathbf{W}_i, \boldsymbol{\gamma}_0)$ is correctly specified. However, this model may be misspecified in practice. We deal with this issue in three ways. First, we suggest the practical guideline of using flexible parametric models for time $T$ and the other potential continuous auxiliary variables to make the correct specification assumption more plausible, or at least to provide a better approximation to the true model for $\pi_j(\mathbf{W}_i)$. This can be achieved by incorporating logarithmic, quadratic and higher order terms, or (finite-dimensional) B-spline terms, where the number of internal knots is fixed and does not depend on sample size $n$. Second, we provide a residual process to formally evaluate the parametric assumption regarding $\pi_j(\mathbf{W}_i, \boldsymbol{\gamma}_0)$ in the next section. Finally, we evaluate the robustness of our estimator against misspecification of $\pi_j(\mathbf{W}_i, \boldsymbol{\gamma}_0)$ in simulation studies.

## 3 Methodology

### 3.1 Estimators

In the ideal situation where the cause of failure is fully observed, that is $C_i$ is available for all $i = 1, 2, \ldots, n$, one can estimate $\boldsymbol{\beta}_0 = (\boldsymbol{\beta}_{0,1}^T, \ldots, \boldsymbol{\beta}_{0,k}^T)^T$ in (2) by maximizing the usual partial likelihood:

$$\begin{aligned}
pl_n(\boldsymbol{\beta}) &= \sum_{j=1}^k \sum_{i=1}^n \int_0^\tau \left\{ \boldsymbol{\beta}_j^T \mathbf{Z}_i - \log\left[\sum_{l=1}^n Y_l(t) e^{\boldsymbol{\beta}_j^T \mathbf{Z}_l}\right] \right\} dN_{ij}(t) \\
&\equiv \sum_{j=1}^k pl_{n,j}(\boldsymbol{\beta}_j).
\end{aligned} \quad (3)$$

If there are no restrictions that the hazards for different causes of failure share the same regression coefficient values, estimation of $\boldsymbol{\beta}_{0,j}$ for any $j = 1, \ldots, k$, can be performed by independently maximizing $pl_{n,j}(\boldsymbol{\beta}_j)$. When some causes of failure are missing, the partial likelihood (3) cannot be evaluated. In this case, the expected log partial likelihood, conditionally on the observed data $\{\mathbf{D}_i\}_{i=1}^n$ is

$$Q_n(\boldsymbol{\beta}) = \sum_{j=1}^{k} \sum_{i=1}^{n} \int_0^\tau \left\{ \boldsymbol{\beta}_j^T \mathbf{Z}_i - \log \left[ \sum_{l=1}^n Y_l(t) e^{\boldsymbol{\beta}_j^T \mathbf{Z}_l} \right] \right\} dE[N_{ij}(t)|\mathbf{D}_i], \quad (4)$$

where

$$E[N_{ij}(t)|\mathbf{D}_i] = [R_i \Delta_{ij} + (1 - R_i)\pi_j(\mathbf{W}_i, \boldsymbol{\gamma}_0)] N_i(t)$$
$$\equiv \tilde{N}_{ij}(t; \boldsymbol{\gamma}_0)$$

since $E(\Delta_{ij}|\mathbf{D}_i) = \pi_j(\mathbf{W}_i, \boldsymbol{\gamma}_0)$ if $R_i = 0$. A pseudo-partial-likelihood for $\boldsymbol{\beta}$ can be constructed by replacing the unknown parameters $\boldsymbol{\gamma}_0$ in the expected log partial likelihood (4) with a consistent estimator $\hat{\boldsymbol{\gamma}}_n$. Therefore, under the missing at random assumption, the first stage of the analysis is to estimate $\boldsymbol{\gamma}_0$ by maximum likelihood based on the data with an observed cause of failure (complete cases), assuming for example a multinomial logit model. It has to be noted that this first stage of the analysis is identical to the first stage of the multiple-imputation approach by Lu and Tsiatis (2001). However, unlike Lu and Tsiatis (2001), in the second stage of the analysis we do not utilize simulation-based imputations and, therefore, we avoid the additional variability due to the finite number of imputations (Wang and Robins 1998). For the second stage of the analysis, we construct the estimating functions given $\hat{\boldsymbol{\gamma}}_n$ as follows

$$\mathbf{G}_{n,j}(\boldsymbol{\beta}_j; \hat{\boldsymbol{\gamma}}_n) = \frac{1}{n} \sum_{i=1}^n \int_0^\tau [\mathbf{Z}_i - E_n(t; \boldsymbol{\beta}_j)] d\tilde{N}_{ij}(t; \hat{\boldsymbol{\gamma}}_n), \quad j = 1, \ldots, k,$$

where

$$E_n(t, \boldsymbol{\beta}_j) = \frac{\sum_{i=1}^n \mathbf{Z}_i Y_i(t) \exp(\boldsymbol{\beta}_j^T \mathbf{Z}_i)}{\sum_{i=1}^n Y_i(t) \exp(\boldsymbol{\beta}_j^T \mathbf{Z}_i)}.$$

The second stage of the analysis is to get the estimators $\hat{\boldsymbol{\beta}}_{n,j}$ as the solutions to the equations $\mathbf{G}_{n,j}(\boldsymbol{\beta}_j; \hat{\boldsymbol{\gamma}}_n) = \mathbf{0}$ for $j = 1, \ldots, k$. Computation can be easily implemented using the `coxph` function in the R package `survival`, as illustrated in the Electronic Supplementary Material. However, computation of standard errors requires bootstrap methods as we have not implemented the standard error estimators for general use in the R software yet.

The parametric assumption on the models for $\pi_j(\mathbf{W}_i, \boldsymbol{\gamma}_0)$, $j = 1, \ldots, k$, can be evaluated using the cumulative residual processes

$$E\{R_i[N_{ij}(t) - \pi_j(\mathbf{W}_i, \boldsymbol{\gamma}_0) N_i(t)]\}, \quad t \in [0, \tau], \quad j = 1, \ldots, k,$$

which can be estimated by

$$\frac{1}{n}\sum_{i=1}^{n} R_i[N_{ij}(t) - \pi_j(\mathbf{W}_i, \hat{\boldsymbol{\gamma}}_n)N_i(t)], \quad t \in [0, \tau], \quad j = 1, \ldots, k.$$

Under the null hypothesis of a correctly specified model, the cumulative residual process is equal to 0 for all $t \in [0, \tau]$. A formal goodness of fit test can be performed using a simulation approach similar to that proposed by Pan and Lin (2005). Additionally, a graphical evaluation of goodness of fit can be performed by plotting the simultaneous confidence band for the residual process around the line $f(t) = 0$ and examining whether the observed residual process falls outside the region formed by the confidence band. The latter provides strong evidence for the violation of the correct specification assumption for the model $\pi_j(\mathbf{W}_i, \boldsymbol{\gamma}_0)$. Further details on this goodness of fit evaluation approach can be found in Bakoyannis et al. (2019). This goodness of fit approach is illustrated in Sect. 5.

The cumulative baseline cause-specific hazard functions can be estimated using the Breslow-type estimator

$$\hat{\Lambda}_{n,j}(t) = \int_0^t \frac{\sum_{i=1}^n d\tilde{N}_{ij}(s; \hat{\boldsymbol{\gamma}}_n)}{\sum_{i=1}^n Y_i(s)e^{\hat{\boldsymbol{\beta}}_{n,j}^T \mathbf{Z}_i}}, \quad j = 1, \ldots, k, \quad t \in [0, \tau].$$

Natural estimators of the covariate-specific cumulative incidence functions for $\mathbf{Z} = \mathbf{z}_0$ are given by

$$\hat{F}_{n,j}(t; \mathbf{z}_0) = \int_0^t \exp\left[-\sum_{l=1}^k \hat{\Lambda}_{n,l}(s-; \mathbf{z}_0)\right] d\hat{\Lambda}_{n,j}(s; \mathbf{z}_0), \quad j = 1, \ldots, k, \quad t \in [0, \tau],$$

where $\hat{\Lambda}_{n,j}(t; \mathbf{z}_0) = \hat{\Lambda}_{n,j}(t) \exp(\hat{\boldsymbol{\beta}}_{n,j}^T \mathbf{z}_0)$ for all $j = 1, \ldots, k$ and $t \in [0, \tau]$.

Although we have only considered time-independent covariates here, the proposed estimator for the regression parameter and its properties, provided in the Sect. 3.2, are also valid for the case of time-dependent covariates, provided that these covariates are right-continuous with left-hand limits and of bounded variation. However, inference for the baseline cumulative cause-specific hazards and the covariate-specific cumulative incidence functions with internal time-dependent covariates is trickier and requires explicit modeling of the covariate processes (Cortese and Andersen 2010).

### 3.2 Asymptotic properties

Before providing the regularity conditions assumed here, we define the negative of the second derivative of the true log partial likelihood function as

$$\mathbf{H}_j(\boldsymbol{\beta}_j) = \int_0^\tau \left(\frac{E[\mathbf{Z}^{\otimes 2} Y(t)e^{\boldsymbol{\beta}_j^T \mathbf{Z}}]}{E[Y(t)e^{\boldsymbol{\beta}_j^T \mathbf{Z}}]} - \left\{\frac{E[\mathbf{Z}Y(t)e^{\boldsymbol{\beta}_j^T \mathbf{Z}}]}{E[Y(t)e^{\boldsymbol{\beta}_j^T \mathbf{Z}}]}\right\}^{\otimes 2}\right) E[d\tilde{N}_j(t; \boldsymbol{\gamma}_0)],$$

for $j = 1, \ldots, k$. The asymptotic properties of the proposed estimators are studied under the following regularity conditions:

C1. The follow-up interval is $[0, \tau]$, with $\tau < \infty$ and $\Lambda_{0,j}(t)$ is a non-decreasing continuous function with $\Lambda_{0,j}(\tau) < \infty$ for each $j = 1, \ldots, k$. Additionally, $E[Y(\tau)|\mathbf{Z}] > 0$ almost surely.
C2. $\boldsymbol{\beta}_{0,j} \in \mathcal{B}_j \subset \mathbb{R}^{p_j}$ where $\mathcal{B}_j$ is a bounded and convex set for all $j = 1, \ldots, k$ and $\boldsymbol{\beta}_{0,j}$ is in the interior of $\mathcal{B}_j$.
C3. The inverse $g$ of the link function for the parametric cause of failure probability model $\pi_j(\mathbf{W}, \boldsymbol{\gamma}_0)$, $j = 1, \ldots, k$, has a continuous derivative $\dot{g}$ with respect to $\boldsymbol{\gamma}_0$ on compact sets. Also, the corresponding parameter space $\Gamma$ for $\boldsymbol{\gamma}_0$ is a bounded subset of $\mathbb{R}^p$.
C4. The score function $U(\boldsymbol{\gamma})$ for the model for the true failure type $C$ is Lipschitz continuous in $\boldsymbol{\gamma}$ and the estimator $\hat{\boldsymbol{\gamma}}_n$ is almost surely consistent and asymptotically linear, i.e. $\sqrt{n}(\hat{\boldsymbol{\gamma}}_n - \boldsymbol{\gamma}_0) = n^{-1/2} \sum_{i=1}^{n} \boldsymbol{\omega}_i + o_p(1)$, with the influence function $\boldsymbol{\omega}_i$ satisfying $E(\boldsymbol{\omega}_i) = \mathbf{0}$ and $E\|\boldsymbol{\omega}_i\|^2 < \infty$ for all $i = 1, 2, \ldots, n$. Additionally, the plug-in estimators $\hat{\boldsymbol{\omega}}_i$ for $\boldsymbol{\omega}_i$ satisfy $n^{-1} \sum_{i=1}^{n} \|\hat{\boldsymbol{\omega}}_i - \boldsymbol{\omega}_i\|^2 = o_p(1)$.
C5. The covariate vector $\mathbf{Z}$ and auxiliary covariate vector $\mathbf{A}$ are bounded in the sense that there exists a constant $K \in (0, \infty)$ such that $P(\|\mathbf{Z}\| \vee \|\mathbf{A}\| \leq K) = 1$.
C6. The true Hessian matrix $-\mathbf{H}_j(\boldsymbol{\beta}_j)$ is a negative definite matrix for all $j = 1, \ldots, k$.

**Remark 1** Conditions C3 and C4 are automatically satisfied if the model for $\pi_j(\mathbf{W}, \boldsymbol{\gamma}_0)$ is a correctly specified binary or multinomial logit model with model parameters estimated through maximum likelihood.

The asymptotic properties of the proposed estimators are provided in the following theorems. The proofs of these theorems are provided in the Electronic Supplementary Material.

**Theorem 1** *Given the assumptions stated in Sect. 2 and the regularity conditions C1–C6,*

$$\sum_{j=1}^{k} \left( \|\hat{\boldsymbol{\beta}}_{n,j} - \boldsymbol{\beta}_{0,j}\| + \|\hat{\Lambda}_{n,j}(t) - \Lambda_{0,j}(t)\|_\infty \right) \overset{as*}{\to} 0$$

*where* $\|f(t)\|_\infty = \sup_{t \in [0,\tau]} |f(t)|$.

**Remark 2** Based on this consistency result it is easy to argue that $\sum_{j=1}^{k} \|\hat{\Lambda}_{n,j}(t; \mathbf{z}_0) - \Lambda_{0,j}(t; \mathbf{z}_0)\|_\infty \overset{as*}{\to} 0$ for any $\mathbf{z}_0$ in the (bounded) covariate space. This fact along with a continuity result from the Duhamel equation (Andersen et al. 1993) can be used to show that $\sum_{j=1}^{k} \|\hat{F}_{n,j}(t; \mathbf{z}_0) - F_{0,j}(t; \mathbf{z}_0)\|_\infty \overset{as*}{\to} 0$ for any $\mathbf{z}_0$ in the (bounded) covariate space, since $\hat{F}_{n,j}(t; \mathbf{z}_0)$, $j = 1, \ldots, k$, are elements of a product integral matrix (Andersen et al. 1993).

Before providing the theorem for the asymptotic distribution of the finite-dimensional parameter estimator we define some useful quantities. Define

$$\psi_{ij} = \mathbf{H}_j^{-1}(\boldsymbol{\beta}_{0,j}) \int_0^\tau [\mathbf{Z}_i - E(t, \boldsymbol{\beta}_{0,j})] d\tilde{M}_{ij}(t; \boldsymbol{\beta}_{0,j}, \boldsymbol{\gamma}_0)$$

for $i = 1, \ldots, n$ and $j = 1, \ldots, k$, where

$$E(t, \boldsymbol{\beta}_{0,j}) = \frac{E[\mathbf{Z}Y(t)e^{\boldsymbol{\beta}_{0,j}^T \mathbf{Z}}]}{E[Y(t)e^{\boldsymbol{\beta}_{0,j}^T \mathbf{Z}}]}$$

and $\tilde{M}_{ij}(t; \boldsymbol{\beta}_{0,j}, \boldsymbol{\gamma}_0) = \tilde{N}_{ij}(t; \boldsymbol{\gamma}_0) - \int_0^t Y_i(s) \exp(\boldsymbol{\beta}_{0,j}^T \mathbf{Z}_i) d\Lambda_{0,j}(s)$, with

$$\Lambda_{0,j}(t) = \int_0^t \frac{E[d\tilde{N}_j(s; \boldsymbol{\gamma}_0)]}{E[Y(s)e^{\boldsymbol{\beta}_{0,j}^T \mathbf{Z}}]}.$$

Finally, define the non-random quantity

$$\mathbf{R}_j = \mathbf{H}_j^{-1}(\boldsymbol{\beta}_{0,j}) \left( E \left\{ (1-R) \int_0^\tau [\mathbf{Z} - E(t, \boldsymbol{\beta}_{0,j})] dN(t) \dot{\pi}_j(\mathbf{W}, \boldsymbol{\gamma}_0)^T \right\} \right)$$

where $\dot{\pi}_j(\mathbf{W}, \boldsymbol{\gamma}_0) = \partial[\pi_j(\mathbf{W}, \boldsymbol{\gamma})](\partial \boldsymbol{\gamma})^{-1}|_{\boldsymbol{\gamma} = \boldsymbol{\gamma}_0}$ and $\boldsymbol{\omega}_i = \mathcal{I}^{-1}(\boldsymbol{\gamma}_0) \mathbf{U}_i(\boldsymbol{\gamma}_0)$ is the influence function for $\hat{\boldsymbol{\gamma}}_n$, with $\mathcal{I}(\boldsymbol{\gamma}_0)$ being the true Fisher information about $\boldsymbol{\gamma}_0$ and $\mathbf{U}_i(\boldsymbol{\gamma}_0)$ the individual score function for the $i$th subject. The following theorem provides the basis for performing statistical inference regarding the finite-dimensional parameter.

**Theorem 2** *Given the assumptions stated in Sect.* 2 *and the regularity conditions C1–C6,*

$$\sqrt{n}(\hat{\boldsymbol{\beta}}_{n,j} - \boldsymbol{\beta}_{0,j}) = \frac{1}{\sqrt{n}} \sum_{i=1}^n (\boldsymbol{\psi}_{ij} + \mathbf{R}_j \boldsymbol{\omega}_i) + o_p(1),$$

*and therefore* $\sqrt{n}(\hat{\boldsymbol{\beta}}_{n,j} - \boldsymbol{\beta}_{0,j})$ *converges in distribution to a mean-zero Gaussian random vector with covariance matrix* $\boldsymbol{\Sigma}_j = E(\boldsymbol{\psi}_j + \mathbf{R}_j \boldsymbol{\omega})^{\otimes 2}$ *that is bounded for all* $j = 1, \ldots, k$.

**Remark 3** The covariance matrix $\Sigma_j$ can be consistently (in probability) estimated by

$$\hat{\boldsymbol{\Sigma}}_j = \frac{1}{n} \sum_{i=1}^n (\hat{\boldsymbol{\psi}}_{ij} + \hat{\mathbf{R}}_j \hat{\boldsymbol{\omega}}_i)^{\otimes 2},$$

where the estimated components of the influence functions in $\hat{\boldsymbol{\Sigma}}_j$ are the empirical estimates of the influence function components defined above, with the unknown parameters being replaced by their consistent estimates and the expectations by sample averages. Explicit formulas for the estimated influence functions are provided in the Electronic Supplementary Material.

Before stating the theorem for the asymptotic distribution of $\hat{\Lambda}_{n,j}$ we define the influence functions

$$\phi_{ij}(t) = \int_0^t \frac{d\tilde{M}_{ij}(s;\boldsymbol{\beta}_{0,j},\boldsymbol{\gamma}_0)}{E[Y(s)e^{\boldsymbol{\beta}_{0,j}^T \mathbf{Z}}]} - (\boldsymbol{\psi}_{ij} + \mathbf{R}_j \boldsymbol{\omega}_i)^T \int_0^t E(s,\boldsymbol{\beta}_{0,j}) d\Lambda_{0,j}(s)$$

and the non-random function

$$\mathbf{R}_j^\star(t) = E\left\{(1-R)\dot{\boldsymbol{\pi}}_j(\mathbf{W},\boldsymbol{\gamma}_0) \int_0^t \frac{dN(s)}{E[Y(s)e^{\boldsymbol{\beta}_{0,j}^T \mathbf{Z}}]}\right\}^T.$$

**Theorem 3** *Given the assumptions stated in Sect. 2 and the regularity conditions C1–C6,*

$$\sqrt{n}\left[\hat{\Lambda}_{n,j}(t) - \Lambda_{0,j}(t)\right] = \frac{1}{\sqrt{n}}\sum_{i=1}^n \left[\phi_{ij}(t) + \mathbf{R}_j^\star(t)\boldsymbol{\omega}_i\right] + o_p(1), \quad (5)$$

*and the influence functions $\phi_{ij}(t) + \mathbf{R}_j^\star(t)\boldsymbol{\omega}_i$ belong to a Donsker class indexed by $t \in [0,\tau]$. Therefore, (5) converges weakly to a tight mean-zero Gaussian process in the space $D[0,\tau]$ of right-continuous functions with left-hand limits, defined on $[0,\tau]$, for all $j = 1,\ldots,k$, with covariance function $E[\phi_j(t)+\mathbf{R}_j^\star(t)\boldsymbol{\omega}][\phi_j(s)+\mathbf{R}_j^\star(s)\boldsymbol{\omega}]$, for $t,s \in [0,\tau]$. Additionally, $\hat{W}_{n,j}(t) = n^{-1/2}\sum_{i=1}^n [\hat{\phi}_{ij}(t)+\hat{\mathbf{R}}_j^\star(t)\hat{\boldsymbol{\omega}}_i]\xi_i$, where $\{\xi_i\}_{i=1}^n$ are standard normal variables independent of the data, converges weakly (conditionally on the data) to the same limiting process as $W_{n,j}(t) = n^{-1/2}\sum_{i=1}^n [\phi_{ij}(t) + \mathbf{R}_j^\star(t)\boldsymbol{\omega}_i]$ (unconditionally).*

*Remark 4* The covariance function can be uniformly consistently (in probability) estimated by

$$\frac{1}{n}\sum_{i=1}^n [\hat{\phi}_{ij}(t) + \hat{\mathbf{R}}_j^\star(t)\hat{\boldsymbol{\omega}}_i][\hat{\phi}_{ij}(s) + \hat{\mathbf{R}}_j^\star(s)\hat{\boldsymbol{\omega}}_i].$$

where $\hat{\phi}_{ij}(t)$, $\hat{\mathbf{R}}_j^\star(t)$ and $\hat{\boldsymbol{\omega}}_i$ are the empirical estimates of the corresponding true functions with the unknown parameters being replaced by their consistent estimates and the expectations by sample averages.

The asymptotic result of Theorem 3 can be straightforwardly used for the construction of $1-\alpha$ pointwise confidence intervals. For the construction of simultaneous confidence bands we use a similar approach to that proposed by Spiekerman and Lin (1998). Consider the process $\sqrt{n}q_j^\Lambda(t)\{g[\hat{\Lambda}_{n,j}(t)] - g[\Lambda_{0,j}(t)]\}$, where $g$ is a known continuously differentiable transformation with nonzero derivative and $q_j^\Lambda$ is a weight function that converges uniformly in probability to a nonnegative bounded function on $[t_1,t_2]$, with $0 \leq t_1 \leq t_2 < \tau$. The transformation ensures that the limits of the confidence band lie within the range of $\Lambda_{0,j}(t)$. For example one can use the transformation $g(x) = \log(x)$ (Lin et al. 1994). The weight function $q_j^\Lambda$, which is useful in reducing the width of the band, can be set equal to $\hat{\Lambda}_{n,j}(t)/\hat{\sigma}_{\Lambda_j}(t)$ with $\hat{\sigma}_{\Lambda_j}(t) = \{n^{-1}\sum_{i=1}^n [\hat{\phi}_{ij}(t) + \hat{\mathbf{R}}_j^\star(t)\hat{\boldsymbol{\omega}}_i]^2\}^{1/2}$, which is the standard error estimate of $W_{n,j}(t)$. This results in the equal precision band (Nair 1984). Another choice for the weight function is $\hat{\Lambda}_{n,j}(t)/[1 + \hat{\sigma}_{\Lambda_j}^2(t)]$ and this results in the Hall–Wellner band (Hall and Wellner 1980). Using the functional delta method it can be shown that the process $\sqrt{n}q_j^\Lambda(t)\{g[\hat{\Lambda}_{n,j}(t)] - g[\Lambda_{0,j}(t)]\}$ is asymptotically equivalent to $B_{n,j}(t) = q_j^\Lambda(t)\dot{g}[\hat{\Lambda}_{n,j}(t)]W_{n,j}(t)$. Furthermore, Theorem 3 ensures that $B_{n,j}(t)$ is asymptotically equivalent to $\hat{B}_{n,j}(t) = q_j^\Lambda(t)\dot{g}\{\hat{\Lambda}_{n,j}(t)\}\hat{W}_{n,j}(t)$. Hence, a $1 - \alpha$ confidence band can be constructed as

$$g^{-1}\left[g\{\hat{\Lambda}_{n,j}(t)\} \pm \frac{c_a}{\sqrt{n}q_j^\Lambda(t)}\right] \quad t \in [t_1, t_2],$$

where $c_\alpha$ is the $1 - a$ quantile of the distribution of $\sup_{t \in [t_1,t_2]} |\hat{B}_{n,j}(t)|$ which can be estimated by the $1 - \alpha$ percentile of the distribution of a large number of simulation realizations of $\sup_{t \in [t_1,t_2]} |\hat{B}_{n,j}(t)|$ (Spiekerman and Lin 1998). Each simulated realization of $\sup_{t \in [t_1,t_2]} |\hat{B}_{n,j}(t)|$ is calculated based on a set of draws of $\{\xi_i\}_{i=1}^n$ values from the standard normal distribution.

**Remark 5** The region of the confidence band $[t_1, t_2]$ typically ranges from the minimum to the maximum observed times of failure from the $j$th type. In order to prevent the effect of the instability in the tails of the cumulative baseline cause-specific hazards estimator, the range can be restricted to $[s_1, s_2]$, where $s_l, l = 1, 2$, can be set equal to the solutions of $c_l = \hat{\sigma}_{\Lambda_j}^2(s_l)/[1 + \hat{\sigma}_{\Lambda_j}^2(s_l)]$, with $\{c_1, c_2\}$ being equal to $\{0.1, 0.9\}$ or $\{0.05, 0.95\}$ (Nair 1984; Yin and Cai 2004).

**Remark 6** It can be also easily shown that $\sqrt{n}[\hat{\Lambda}_{n,j}(t; \mathbf{z}_0) - \Lambda_{0,j}(t; \mathbf{z}_0)]$ is an asymptotically linear estimator with influence functions $\phi_{ij}^\Lambda(t; \mathbf{z}_0) = [\mathbf{z}_0^T(\boldsymbol{\psi}_{ij} + \mathbf{R}_j\boldsymbol{\omega}_i)\Lambda_{0,j}(t) + \phi_{ij}(t) + \mathbf{R}_j^\star(t)\boldsymbol{\omega}_i]\exp(\boldsymbol{\beta}_{0,j}^T\mathbf{z}_0)$ for $j = 1, \ldots, k$ and $t \in [0, \tau]$. The Donsker property of the class $\{\phi_j^\Lambda(t; \mathbf{z}_0) : t \in [0, \tau]\}$, for every $j = 1, \ldots, k$ and $\mathbf{z}_0$ in the bounded covariate space follows from the fact that it is formed by a sum of functions that belong to Donsker classes, which are multiplied by fixed functions. Pointwise $1 - \alpha$ confidence intervals and simultaneous confidence bands can be similarly constructed based on the estimated influence functions $\hat{\phi}_{ij}^\Lambda(t; \mathbf{z}_0)$.

The following theorem describes the asymptotic properties of the plug-in estimators of the covariate-specific cumulative incidence functions.

**Theorem 4** *Given the assumptions stated in Sect. 2 and the regularity conditions C1–C6,*

$$\sqrt{n}\left[\hat{F}_{n,j}(t; \mathbf{z}_0) - F_{0,j}(t; \mathbf{z}_0)\right] = \frac{1}{\sqrt{n}}\sum_{i=1}^n \phi_{ij}^F(t; \mathbf{z}_0) + o_p(1), \quad (6)$$

*where*

$$\phi_{ij}^F(t; \mathbf{z}_0) = \int_0^t \exp\left[-\sum_{l=1}^k \Lambda_{0,l}(s-; \mathbf{z}_0)\right] d\phi_{ij}^\Lambda(s; \mathbf{z}_0)$$
$$- \int_0^t \left[\sum_{l=1}^k \phi_{il}^\Lambda(s-; \mathbf{z}_0)\right] \exp\left[-\sum_{l=1}^k \Lambda_{0,l}(s-; \mathbf{z}_0)\right] d\Lambda_{0,j}(s; \mathbf{z}_0)$$

*and the influence functions* $\phi_{ij}^F(t; \mathbf{z}_0)$ *for* $i = 1, \ldots, n$ *and* $j = 1, \ldots, k$ *belong to a Donsker class indexed by* $t \in [0, \tau]$. *Therefore,* (6) *converges weakly to a tight mean-zero Gaussian process in* $D[0, \tau]$, *for all* $j = 1, \ldots, k$, *with covariance function* $E[\phi_j^F(t; \mathbf{z}_0)\phi_j^F(s; \mathbf{z}_0)]$, *for* $t, s \in [0, \tau]$.

**Remark 7** The covariance function can be uniformly consistently (in probability) estimated by $n^{-1} \sum_{i=1}^n \hat{\phi}_{ij}^F(t; \mathbf{z}_0)\hat{\phi}_{ij}^F(s; \mathbf{z}_0)$, where the empirical influence function $\hat{\phi}_{ij}^F(s; \mathbf{z}_0)$ can be similarly calculated as described above. Moreover, the asymptotic (conditional on the data) distribution of

$$\hat{W}_{n,j}^F(t; \mathbf{z}_0) = \frac{1}{\sqrt{n}} \sum_{i=1}^n \hat{\phi}_{ij}^F(t; \mathbf{z}_0)\xi_i,$$

where $\{\xi_i\}_{i=1}^n$ are standard normal variables independent of the data, is the same as the (unconditional) asymptotic distribution of

$$W_{n,j}^F(t; \mathbf{z}_0) = n^{-1/2} \sum_{i=1}^n \phi_{ij}^F(t; \mathbf{z}_0).$$

**Remark 8** Theorem 4 can be used for the construction of $1 - \alpha$ pointwise confidence intervals for $F_{0,j}(t; \mathbf{z}_0)$. Construction of simultaneous confidence bands can be performed as described for $\Lambda_{0,j}(t)$ and in a similar fashion as that in Cheng et al. (1998), using the process $\hat{B}_{n,j}^F(t; \mathbf{z}_0) = q_j^F(t; \mathbf{z}_0)\dot{g}[\hat{F}_{n,j}(t; \mathbf{z}_0)]\hat{W}_{n,j}^F(t; \mathbf{z}_0)$. In this case the transformation $g(x)$ can be set equal to $\log[-\log(x)]$, and the weight function $q_j^F(t; \mathbf{z}_0)$ to $\hat{F}_{n,j}(t; \mathbf{z}_0)\log[\hat{F}_{n,j}(t; \mathbf{z}_0)]/\hat{\sigma}_{F_j}(t; \mathbf{z}_0)$, with

$$\hat{\sigma}_{F_j}(t; \mathbf{z}_0) = \left\{n^{-1} \sum_{i=1}^n [\hat{\phi}_{ij}^F(t; \mathbf{z}_0)]^2\right\}^{1/2},$$

which is the standard error estimate of $W_{n,j}^F(t; \mathbf{z}_0)$. This weight leads to an equal-precision-type confidence band (Nair 1984). Alternatively, $q_j^F(t; \mathbf{z}_0)$ can be set equal to

$$\hat{F}_{n,j}(t; \mathbf{z}_0)\log[\hat{F}_{n,j}(t; \mathbf{z}_0)]/[1 + \hat{\sigma}_{F_j}^2(t; \mathbf{z}_0)],$$

which yields a Hall–Wellner type confidence band (Hall and Wellner 1980).

## 4 Simulation studies

To evaluate the finite sample performance of the proposed estimator, we conducted a series of simulation studies. We used similar simulation settings to those used in Hyun et al. (2012). Specifically, we considered a cohort study with an observation interval [0, 2], two causes of failure, and two covariates $\mathbf{Z} = (Z_1, Z_2)^T$, where $Z_1$ was generated from $U(0, 1)$ and $Z_2$ from the Bernoulli(0.5) distribution. Additionally, we considered an independent random right-censoring variable simulated from an exponential distribution with a rate equal to 0.4. Event time for cause of failure 1 was generated from the exponential distribution with hazard $\lambda_{0,1}(t; \mathbf{Z}) = \exp(\beta_1 Z_1)$, where $\beta_1 = -0.5$. Event time for cause of failure 2 was generated either from a Gompertz distribution with a rate $\lambda_{0,2}(t; \mathbf{Z}) = \exp[-\beta_2(Z_2 + 1) + \nu t]$ where $(\beta_2, \nu) = (0.5, 0.2)$ (scenario 1), or from a Weibull distribution with a hazard function $\eta \lambda^\eta \exp(\beta_3 Z_2) t^{\eta-1}$ where $(\lambda, \beta_3) = (0.5, -0.5)$ and $\eta = 0.5$ (scenario 2), $\eta = 2$ (scenario 3), or $\eta = 0.1$ (scenario 4). The implied model for $\pi_1(\mathbf{W}, \boldsymbol{\gamma})$, the probability of the cause of failure 1 with $\mathbf{W} = (T, \mathbf{Z})$, has the form

$$\text{logit}[\pi_1(\mathbf{W}, \boldsymbol{\gamma})] = \gamma_0 + \gamma_1 T + \gamma_2 Z_1 + \gamma_3 Z_2$$

with $(\gamma_0, \gamma_1, \gamma_2, \gamma_3) = (\beta_2, -\nu, \beta_1, \beta_2) = (0.5, -0.2, -0.5, 0.5)$ under scenario 1 and

$$\text{logit}[\pi_1(\mathbf{W}, \boldsymbol{\gamma})] = \gamma_0 + \gamma_1 \log(T) + \gamma_2 Z_1 + \gamma_3 Z_2$$

with

$$(\gamma_0, \gamma_1, \gamma_2, \gamma_3) = (-\log(\eta) + \lambda \eta, -(\eta - 1), \beta_1, \lambda \eta [\exp(\beta_3) - 1])$$

under scenarios 2-4. For scenario 2, $(\gamma_0, \gamma_1, \gamma_2, \gamma_3) \approx (0.94, 0.5, -0.5, -0.10)$, while for scenarios 3 and 4 $(\gamma_0, \gamma_1, \gamma_2, \gamma_3)$ was equal to $(0.31, -1, -0.5, -0.39)$ and $(2.35, 0.9, -0.5, -0.02)$, respectively. This simulation setup resulted on average in 25.6% right-censored observations and 59.4% failures from cause 1 and 40.6% failures from cause 2, under scenario 1. The corresponding figures for scenarios 2-4 were $25.1\% - 54.1\% - 45.9$, $31.6\% - 77.7\% - 22.3\%$, and $20.0\% - 38.3\% - 61.7\%$, respectively. The average ranges of failure time in scenarios 1-4 were $0.004 - 1.901$, $< 0.001 - 1.894$, $0.006 - 1.932$, and $< 0.001 - 1.874$. For the probability of an observed cause of failure $P(R = 1 | \Delta_i = 1, \mathbf{W}) \equiv p(\mathbf{W}, \boldsymbol{\theta})$ (i.e. 1 - probability of missingness) we considered a model of the form

$$\text{logit}[p(\mathbf{W}, \boldsymbol{\theta})] = \theta_0 + \theta_1 T + \theta_2 Z_1 + \theta_3 Z_2.$$

In our simulations we considered $\boldsymbol{\theta} = (0.7, 1, -1, 1)^T$, $\boldsymbol{\theta} = (-0.2, 1, -1, 1)^T$, or $\boldsymbol{\theta} = (-0.8, 1, -1, 1)^T$ which resulted in 25.2%, 43.5% and 56.4% missingness on average under scenario 1, 27.1%, 45.5% and 58.6% missingness under scenario 2, 23.1%, 40.4% and 53.6% missingness under scenario 3, and 30.2%, 49.3% and 62.2% missingness under scenario 4.

For each scenario we simulated 1000 datasets and evaluated the performance of the proposed MPPLE, the AIPW estimator (Gao and Tsiatis 2005; Hyun et al. 2012), and the multiple imputation (MI) estimator with 5 imputations (Lu and Tsiatis 2001), for estimating $\beta_1$. For the AIPW estimator, we used the correctly specified model $p(\mathbf{W}, \boldsymbol{\theta})$ for the probability of an observed cause of failure in all cases to guarantee the estimation consistency due to its double robustness property. For the probability of $C = 1$ given $\{\Delta = 1\}$ and $\mathbf{W} = (T, \mathbf{Z})$, all analyses assumed the model $\text{logit}[\pi_1(\mathbf{W}, \boldsymbol{\gamma})] = \gamma_0 + \gamma_1 T + \gamma_2 Z_1 + \gamma_3 Z_2$. Therefore, the assumed model for $\pi_1(\mathbf{W}, \boldsymbol{\gamma})$ was correctly specified in scenario 1, but misspecified in scenarios 2-4. For standard error estimation we used the proposed closed-form estimators provided in Sect. 3.2 for the proposed MPPLE, while for the AIPW and the MI estimators we used bootstrap based on 100 replications. We also evaluated the performance of our estimators for the infinite-dimensional parameters. The simultaneous 95% confidence bands for these parameters were constructed based on 1000 simulation realizations of sets $\{\xi_i\}_{i=1}^n$, from the standard normal distribution. The domain limits for the confidence bands were calculated based on $\{c_1, c_2\} = \{0.1, 0.9\}$, as described in the preceding section. Note that since the AIPW approach (Gao and Tsiatis 2005; Hyun et al. 2012) and the MI estimator (Lu and Tsiatis 2001) did not consider inference about the infinite-dimensional parameters we were not able to provide results from these approaches in the latter set of simulations.

Simulation results for the regression coefficient $\beta_1$ under scenario 1 are presented in Table 1. The MPPLE provides virtually unbiased estimates even under a mispecified model $\pi_1(\mathbf{W}, \boldsymbol{\gamma})$. The average standard error estimates are close to the corresponding Monte Carlo standard deviations of the estimates, with the empirical coverage probabilities being close to the nominal level in all cases. Compared to the AIPW estimator with the correctly specified model for the probability of an observed cause of failure and the MI estimator, our estimator achieves higher efficiency in all cases. The advantage of our estimator over the AIPW estimator in terms of efficiency is substantial in cases with a larger sample size and a larger proportion of missing cause of failure. However, such a pattern was not observed for the case of the MI estimator. Simulation results under scenario 2 (Table 2) are similar. These results indicate the robustness of our estimator against certain misspecification of the parametric model $\pi_1(\mathbf{W}, \boldsymbol{\gamma})$ and, also, its substantially higher efficiency compared to the AIPW estimators in cases with larger sample size and proportion of missing cause of failure. Simulation results under scenarios 3 and 4 with a more pronounced misspecification of the model $\pi_1(\mathbf{W}, \boldsymbol{\gamma})$ (Tables 1 and 2 in the Electronic Supplementary Material) are similar, although the higher efficiency of our estimator compared to the AIPW estimator is less pronounced in these cases. It has to be noted that, under a scenario with baseline hazards of a more complicated form or a considerably longer follow-up period, it is expected that the MPPLE and the multiple imputation estimator would exhibit more bias and lower coverage rates. Simulation results for the infinite-dimensional parameters are presented in Tables 3–8 in the Electronic Supplementary Material. The bias of our estimators is very small even in cases where $\pi_1(\mathbf{W}, \boldsymbol{\gamma})$ is misspecified, the average standard error estimates are close to the corresponding Monte Carlo standard deviations of the estimates, and the empirical coverage probabilities for the pointwise confidence intervals remain close to the nominal level in scenarios 1 and 2. In scenarios 3 and 4, where the

**Table 1** Simulation results for $\beta_1$ under scenario 1 where the model $\pi_1(\mathbf{W}, \boldsymbol{\gamma})$ was correctly specified

| n | $p_m$ (%) | Method | Bias | MCSD | ASE | CP | MSE | RE |
|---|---|---|---|---|---|---|---|---|
| 200 | 25 | Proposed MPPLE | 0.002 | 0.409 | 0.396 | 0.945 | 0.167 | 1.000 |
| | | AIPW | 0.003 | 0.412 | 0.418 | 0.946 | 0.170 | 1.013 |
| | | MI(5) | 0.009 | 0.424 | 0.419 | 0.941 | 0.180 | 1.074 |
| | 44 | Proposed MPPLE | 0.007 | 0.450 | 0.428 | 0.943 | 0.203 | 1.000 |
| | | AIPW | 0.009 | 0.464 | 0.468 | 0.943 | 0.215 | 1.061 |
| | | MI(5) | 0.004 | 0.460 | 0.461 | 0.946 | 0.211 | 1.043 |
| | 56 | Proposed MPPLE | 0.004 | 0.492 | 0.468 | 0.942 | 0.242 | 1.000 |
| | | AIPW | 0.009 | 0.526 | 0.540 | 0.949 | 0.277 | 1.144 |
| | | MI(5) | −0.004 | 0.502 | 0.510 | 0.951 | 0.253 | 1.043 |
| 400 | 25 | Proposed MPPLE | 0.001 | 0.284 | 0.282 | 0.948 | 0.081 | 1.000 |
| | | AIPW | −0.001 | 0.289 | 0.288 | 0.949 | 0.084 | 1.038 |
| | | MI(5) | −0.004 | 0.290 | 0.290 | 0.948 | 0.084 | 1.046 |
| | 44 | Proposed MPPLE | −0.001 | 0.308 | 0.305 | 0.949 | 0.095 | 1.000 |
| | | AIPW | −0.004 | 0.326 | 0.321 | 0.946 | 0.106 | 1.116 |
| | | MI(5) | −0.008 | 0.320 | 0.316 | 0.950 | 0.102 | 1.076 |
| | 56 | Proposed MPPLE | −0.003 | 0.337 | 0.333 | 0.946 | 0.114 | 1.000 |
| | | AIPW | −0.008 | 0.368 | 0.364 | 0.937 | 0.135 | 1.191 |
| | | MI(5) | −0.006 | 0.350 | 0.346 | 0.940 | 0.122 | 1.077 |
| 2000 | 25 | Proposed MPPLE | 0.003 | 0.124 | 0.126 | 0.955 | 0.015 | 1.000 |
| | | AIPW | 0.003 | 0.126 | 0.127 | 0.950 | 0.016 | 1.029 |
| | | MI(5) | 0.003 | 0.127 | 0.127 | 0.955 | 0.016 | 1.045 |
| | 44 | Proposed MPPLE | 0.005 | 0.132 | 0.136 | 0.954 | 0.017 | 1.000 |
| | | AIPW | 0.005 | 0.137 | 0.139 | 0.953 | 0.019 | 1.080 |
| | | MI(5) | 0.003 | 0.139 | 0.138 | 0.950 | 0.019 | 1.119 |
| | 56 | Proposed MPPLE | 0.002 | 0.142 | 0.148 | 0.956 | 0.020 | 1.000 |
| | | AIPW | 0.002 | 0.152 | 0.155 | 0.941 | 0.023 | 1.150 |
| | | MI(5) | 0.003 | 0.153 | 0.150 | 0.946 | 0.023 | 1.164 |

$p_m$, percent of missingness; MCSD, Monte Carlo standard deviation; ASE, average estimated standard error; CP, coverage probability; MSE, mean squared error; RE, variance of the estimator to variance of the proposed MPPLE (relative efficiency); MPPLE, maximum partial pseudolikelihood estimator; AIPW, augmented inverse probability weighting estimator; MI(5), Lu and Tsiatis type B multiple imputation based on 5 imputations model misspecification is more pronounced, empirical coverage probabilities were also close to the nominal level except for the early time point that corresponds to the 10% of the total follow-up time, in some cases. The simultaneous confidence bands have empirical coverage probabilities close to the nominal level under a correctly specified model for $\pi_1(\mathbf{W}, \boldsymbol{\gamma})$. However, the coverage of the confidence bands in cases where the event time is modeled incorrectly, i.e. as $T$ instead of $\log(T)$, is lower than 95%, especially in cases with a large fraction of missingness. The latter

**Table 2** Simulation results for $\beta_1$ under scenario 2 where the model $\pi_1(\mathbf{W}, \boldsymbol{\gamma})$ was misspecified with $\eta = 0.5$

| $n$ | $p_m$ (%) | Method | Bias | MCSD | ASE | CP | MSE | RE |
|---|---|---|---|---|---|---|---|---|
| 200 | 27 | Proposed MPPLE | 0.006 | 0.424 | 0.419 | 0.955 | 0.180 | 1.000 |
| | | AIPW | 0.004 | 0.427 | 0.442 | 0.957 | 0.182 | 1.014 |
| | | MI(5) | 0.001 | 0.445 | 0.446 | 0.956 | 0.198 | 1.100 |
| | 46 | Proposed MPPLE | 0.015 | 0.471 | 0.458 | 0.954 | 0.222 | 1.000 |
| | | AIPW | 0.013 | 0.484 | 0.500 | 0.955 | 0.235 | 1.059 |
| | | MI(5) | −0.007 | 0.487 | 0.495 | 0.948 | 0.237 | 1.071 |
| | 59 | Proposed MPPLE | 0.009 | 0.520 | 0.504 | 0.939 | 0.271 | 1.000 |
| | | AIPW | 0.009 | 0.556 | 0.579 | 0.952 | 0.310 | 1.143 |
| | | MI(5) | −0.010 | 0.536 | 0.553 | 0.951 | 0.287 | 1.061 |
| 400 | 27 | Proposed MPPLE | 0.000 | 0.301 | 0.298 | 0.952 | 0.091 | 1.000 |
| | | AIPW | −0.002 | 0.306 | 0.305 | 0.946 | 0.094 | 1.034 |
| | | MI(5) | −0.004 | 0.312 | 0.306 | 0.943 | 0.097 | 1.070 |
| | 46 | Proposed MPPLE | −0.001 | 0.332 | 0.326 | 0.948 | 0.110 | 1.000 |
| | | AIPW | −0.006 | 0.350 | 0.343 | 0.945 | 0.122 | 1.111 |
| | | MI(5) | −0.007 | 0.348 | 0.337 | 0.933 | 0.121 | 1.098 |
| | 59 | Proposed MPPLE | −0.004 | 0.364 | 0.359 | 0.946 | 0.132 | 1.000 |
| | | AIPW | −0.012 | 0.399 | 0.390 | 0.941 | 0.159 | 1.203 |
| | | MI(5) | −0.004 | 0.381 | 0.372 | 0.940 | 0.145 | 1.100 |
| 2000 | 27 | Proposed MPPLE | 0.006 | 0.130 | 0.133 | 0.960 | 0.017 | 1.000 |
| | | AIPW | 0.004 | 0.132 | 0.134 | 0.953 | 0.017 | 1.035 |
| | | MI(5) | 0.003 | 0.132 | 0.134 | 0.957 | 0.018 | 1.044 |
| | 46 | Proposed MPPLE | 0.006 | 0.141 | 0.145 | 0.955 | 0.020 | 1.000 |
| | | AIPW | 0.005 | 0.146 | 0.149 | 0.952 | 0.021 | 1.084 |
| | | MI(5) | 0.002 | 0.150 | 0.147 | 0.950 | 0.023 | 1.141 |
| | 59 | Proposed MPPLE | 0.005 | 0.152 | 0.159 | 0.958 | 0.023 | 1.000 |
| | | AIPW | 0.003 | 0.163 | 0.167 | 0.957 | 0.027 | 1.150 |
| | | MI(5) | −0.001 | 0.163 | 0.161 | 0.952 | 0.027 | 1.153 |

$p_m$, percent of missingness; MCSD, Monte Carlo standard deviation; ASE, average estimated standard error; CP, coverage probability; MSE, mean squared error; RE, variance of the estimator to variance of the proposed MPPLE (relative efficiency); MPPLE, maximum partial pseudolikelihood estimator; AIPW, augmented inverse probability weighting estimator; MI(5), Lu and Tsiatis type B multiple imputation based on 5 imputations result indicates the importance of evaluating the goodness of fit of the assumed model for $\pi_1(\mathbf{W}, \boldsymbol{\gamma})$ using the cumulative residual process given in Sect. 3.1.

It is worth pointing out that the proposed MPPLE method not only enjoys efficiency advantages compared to the AIPW method and the MI estimator, but is also computationally very fast and robust. These advantages rank the proposed method favorably in practical applications, particularly in for large studies like the EA-IeDEA HIV study which is analyzed in Sect. 5.1.

## 5 Data applications

In this section we apply the proposed methods to analyze data from the EA-IeDEA HIV study and the EORTC bladder cancer trial, which were mentioned in the Introduction section. It has to be noted that standard errors for the MPPLE were computed using the closed-form estimators provided in Sect. 3.2, while for the AIPW estimators we used bootstrap based on 100 replications.

### 5.1 HIV data analysis

In this subsection we apply the proposed methodology to the electronic health record data from the EA-IeDEA study to analyze time from ART initiation to disengagement from HIV care or death. In this analysis, disengagement from care was defined as being alive and without HIV care for two months. The data set we used consisted of 6657 HIV-infected patients on ART. The data of those patients were collected during routine clinic visits which were typically scheduled every 4 weeks. The median (IQR) time between two consecutive actual visits in our data set was 28 (28, 56) days. In total, 346 patients died (reported deaths) and 2929 patients missed a scheduled clinic visit for a period of at least two months (loss to clinic). The remaining 3,382 patients were still in care at the end of the study period and hence were treated as right-censored observations. Due to the significant death under-reporting in sub-Saharan Africa, the 2929 lost to clinic patients included both disengagers from HIV care and deceased individuals whose death was not reported to the clinic. Of those patients, 448 (15.3%) were successfully outreached by clinic workers in order to ascertain their vital status and record whether these patients were disengagers or deceased. Among them, 99 (22.1%) were found to have died, indicating a significant death under-reporting issue. Cause of failure (i.e. disengagement from care or death) was missing for the remaining 84.7% of the patients who were lost to clinic and were not outreached. For these data, we assumed a binary logistic model $\pi_1(\mathbf{W}, \boldsymbol{\gamma}_0)$ for the probability of death among patients who were lost to clinic. In order to analyze the EA-IeDEA data using the proposed methodology we first evaluated the goodness of fit of this logistic model. The covariates considered in $\pi_1(\mathbf{W}, \boldsymbol{\gamma}_0)$ were time since ART initiation, gender, age, and CD4 cell count at ART initiation. Descriptive characteristics of the study sample are presented in Table 3.

The goodness of fit evaluation based on the residual process defined in Sect. 3.1 is presented in Fig. 1. Figure 1a clearly indicates the lack of fit for the model with a linear effect of time since ART initiation, as the residual process is outside the 95% confidence band for the early timepoints. It is evident that the fitted model $\pi_1(\mathbf{W}, \hat{\boldsymbol{\gamma}}_n)$ underestimates the probability of death within about the first 12 months since ART initiation. After 2 years there is a tendency for overestimation of the probability of death. The corresponding goodness of fit test is statistically significant ($p$ value < 0.001) indicating strong evidence for model misspecification. We then considered a model with piecewise linear effect of time with a change in slope at 12 months after ART initiation. This is a reasonable change point from a clinical perspective because the probability of death is expected to decrease dramatically during the first

**Table 3** Descriptive statistics for the EA-IeDEA study sample

|  | Cause of failure | | | |
| --- | --- | --- | --- | --- |
|  | In care ($N = 3382$) n (%) | Disengagement ($N = 349$) n (%) | Death ($N = 445$[a]) n (%) | Missing ($N = 2481$) n (%) |
| Gender | | | | |
| Female | 2300 (68.0) | 210 (60.2) | 254 (57.1) | 1,665 (67.1) |
| Male | 1082 (32.0) | 139 (39.8) | 191 (42.9) | 816 (32.9) |
|  | Median (IQR) | Median (IQR) | Median (IQR) | Median (IQR) |
| Age[b] | 37.9 (31.8, 45.4) | 35.5 (29.7, 41.9) | 37.3 (31.3, 46.0) | 35.4 (29.9, 42.7) |
| CD4[c] | 174 (91, 258) | 145 (69, 222) | 88 (39, 180) | 155 (71, 214) |

[a]Includes 346 reported deaths and 99 unreported deaths which were ascertained through outreach
[b]At ART initiation in years
[c]At ART initiation in cells/μl

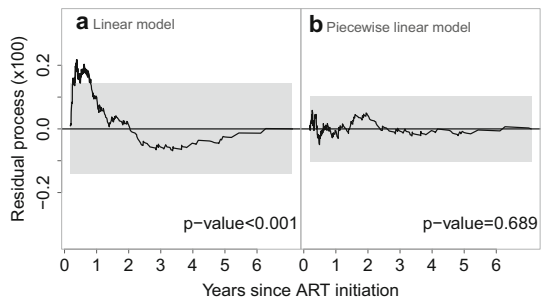

**Fig. 1** Cumulative residual process for the evaluation of the parametric model $\pi_1(W, \gamma_0)$ based on the HIV data along with the 95% goodness-of-fit band (grey area) and the corresponding p value

12 months as a result of ART. After this timepoint the probability of death remains low and approximately constant. The cumulative residual process for this model (Fig. 1b) was close to 0 at all time points and remained within the 95% confidence band under the null hypothesis (p value = 0.689). This piecewise model was used for the analysis of the EA-IeDEA data.

Despite the sample of 6657 patients with a large percent of missing cause of failures, the proposed MPPLE method required only about 33 s for each cause of failure, to compute the regression coefficients and the corresponding standard error estimates. This analysis (Table 4) revealed that males and younger patients have a higher hazard of disengagement from care. Also, patients with a lower CD4 count at ART initiation had a higher hazard of death while in HIV care. The analysis based on the AIPW estimator provided similar results qualitatively, however, unlike the analysis using the proposed MPPLE, the effect of gender was not statistically significant. This is a result of the larger standard error of the AIPW estimator and this is in agreement with our simulation results where our estimator achieved a substantially higher efficiency compared to the AIPW estimator. To illustrate the use of our methodology for risk prediction we depict the predicted cumulative incidence function of disengagement from HIV care and death for a 40-year old male patient with a CD4 cell count of

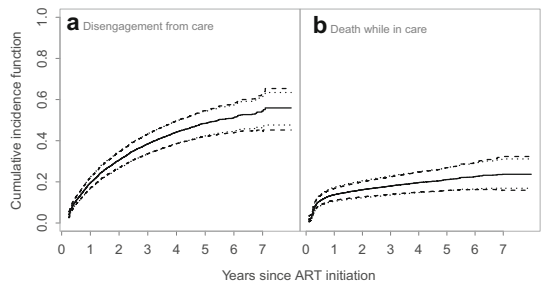

**Fig. 2** Predicted cumulative incidence functions (solid lines) of **a** disengagement from care and **b** death while in HIV care, for a 40-year old male patient with CD4 cell count of 150 cells/μl at ART initiation, along with the 95% simultaneous confidence bands based on equal precision (dotted lines) and Hall–Wellner-type weights (dashed lines)

150 cells/μl at ART initiation, along with the equal-precision and Hall–Wellner-type simultaneous 95% confidence bands, in Fig. 2.

### 5.2 Bladder cancer trial data analysis

In this subsection we analyze a subset of the data from the EORTC bladder cancer clinical trial (Oddens et al. 2013). This trial was conducted to assess whether 1/3 dose of intravesical bacillus Calmette–Guérin (BCG) is inferior to full dose of BCG in treating non-muscle-invasive bladder cancer (NMIBC). The subset of the data we analyze here included 680 intermediate- and high-risk NMIBC patients who underwent transurethral resection and received BCG for one year. Of them, 341 were randomly assigned to the 1/3 dose group and 339 to the full-dose group of the trial. In this analysis, we focus on time to death from bladder cancer (event of interest) or from other causes (competing event). In total, 171 (25.1%) patients died during the study period. Of them, 33 (19.3%) died due to bladder cancer, 115 (67.3%) due to other causes, while the cause of death was missing for 23 (13.4%) of the deceased patients. The covariates considered in this analysis were treatment assignment, age and World Health Organization (WHO) performance status at baseline. Descriptive characteristics of the study sample can be found in Sect. 5 of the Electronic Supplementary Material.

The covariates considered in $\pi_1(\mathbf{W}, \boldsymbol{\gamma}_0)$ were time from randomization to death, treatment assignment, age, and WHO performance status. The goodness of fit evaluation for this model based on the residual process defined in Sect. 3.1 is presented in Figure 1 in the Electronic Supplementary Material. The corresponding goodness of fit test was not statistically significant ($p$ value $= 0.281$) and, thus, there was no evidence for misspecification of $\pi_1(\mathbf{W}, \boldsymbol{\gamma}_0)$. The results of the data analysis regarding the estimated regression coefficients are presented in Table 5 in the Electronic Supplementary Material. The estimated regression coefficient for the effect of assignment to the full-dose BCG group versus the 1/3-dose BCG group on the cause-specific hazard of death from bladder cancer was 0.451 based on the proposed MPPLE estimator and 0.421 based on the AIPW estimator. The corresponding standard error was smaller for the MPPLE (SE $= 0.356$) compared to the AIPW estimator (SE $= 0.372$). However, based on both analyses, the effect of treatment assignment on the cause-specific hazard of death from bladder cancer was not statistically significant. To explicitly test the inferiority null hypothesis that the 1/3 BCG dose assignment is inferior to the full BCG dose assignment, with respect to the cause-specific hazard of death from bladder

**Table 4** Data analysis of the EA-IeDEA study sample

| Covariate | Proposed MPPLE | | | AIPW | | |
|---|---|---|---|---|---|---|
| | $\exp(\hat{\beta}_n)$ | 95% CI | $p$ value | $\exp(\hat{\beta}_n)$ | 95% CI | $p$ value |
| Disengagement from care | | | | | | |
| Sex (male = 1, female = 0) | 1.15 | (1.02, 1.31) | 0.022 | 1.24 | (0.69, 2.23) | 0.462 |
| Age (10 years) | 0.75 | (0.70, 0.80) | < 0.001 | 0.58 | (0.40, 0.85) | 0.004 |
| CD4 (100 cells/μl) | 1.03 | (1.00, 1.06) | 0.094 | 1.17 | (0.97, 1.42) | 0.104 |
| Death while in care | | | | | | |
| Sex (male = 1, female = 0) | 1.24 | (0.96, 1.59) | 0.094 | 1.14 | (0.95, 1.37) | 0.157 |
| Age (10 years) | 1.10 | (0.97, 1.25) | 0.153 | 0.99 | (0.87, 1.13) | 0.926 |
| CD4 (100 cells/μl) | 0.76 | (0.63, 0.91) | 0.003 | 0.78 | (0.68, 0.89) | < 0.001 |

MPPLE: maximum pseudo partial likelihood estimator; AIPW: augmented inverse probability weighting estimator; 95% CI: 95% confidence interval for the cause-specific hazard ratio $\exp(\beta_0)$ cancer, we considered a non-inferiority margin of log(0.85). This non-inferiority margin corresponds to the log hazard ratio for death from bladder cancer in the full-dose BCG group versus the 1/3-dose BCG group. Under this non-inferiority margin and a one-sided Wald test, according to the recommendations for non-inferiority hypothesis testing (Rothmann et al. 2016), the null hypothesis of inferiority of the 1/3 dose assignment compared to the full dose assignment was rejected based on the MPPLE estimator ($p$ value $= 0.042$). However, this null hypothesis could not be rejected at the $\alpha = 0.05$ level based on the AIPW estimator ($p$ value $= 0.059$). To illustrate the use of our methodology for risk prediction we depict the predicted cumulative incidence function of death from bladder cancer and other causes, for a 68-year old patient who is fully active and who was assigned to the 1/3 dose BCG group, along with the equal-precision and Hall–Wellner-type simultaneous 95% confidence bands, in Figure 2 of the Electronic Supplementary Material.

## 6 Concluding remarks

In this article we proposed a computationally efficient MPPLE method for the semiparametric proportional cause-specific hazards model under incompletely observed cause of failure. We propose estimators for both the regression parameters in the proportional cause-specific hazards model and the covariate-specific cumulative incidence functions. To the best of our knowledge, a unified approach for semiparametric inference about both the cause-specific hazard, for evaluating risks factors, and the covariate-specific cumulative incidence function, for risk prediction purposes, is missing in the literature. Our approach utilizes a parametric model for the probability of the cause of failure and imposes a missing at random assumption. The estimators were shown to be strongly consistent and to converge weakly to Gaussian random quantities. Closed-form variance estimators were derived. In addition, we propose methodology for constructing simultaneous confidence bands for the covariate-specific cumulative incidence functions. Simulation studies showed a satisfactory performance of our estimators even under a large fraction of missing causes of failure and under some degree of misspecification of the parametric model for the probability of the cause of failure.

Although the main model of interest is semiparametric, our estimation method depends on the parametric model $\pi_j(\mathbf{W}, \boldsymbol{\gamma}_0)$ for the probability of the cause of failure. Essentially, this model is used to calculate the expected log partial likelihood contribution for the missing cases. The main reason for adopting such a parametric model was to allow the incorporation of auxiliary covariates that are typically important in practice in order to make the MAR assumption plausible. Additionally, this choice led to an increased computational and statistical efficiency of our estimator. It has to be noted that the true model $\pi_j(\mathbf{W}, \boldsymbol{\gamma}_0)$ is induced by the propotional cause-specific specific hazards model assumption and the baseline hazards. Even though correct specification of the model $\pi_j(\mathbf{W}, \boldsymbol{\gamma}_0)$ is a sufficient condition for consistency, our estimator was shown to be robust against some degree of misspecification in the simulation studies. However, the coverage probability of the simultaneous confidence bands was lower than the nominal level when $\pi_j(\mathbf{W}, \boldsymbol{\gamma}_0)$ was misspecified, as a result of bias in the infinite-dimensional parameter estimates. Of course, simulation scenar-ios with more pronounced misspecification are expected to lead to more bias and lower coverage rates. For this reason, we suggest the practical guideline of using flexible parametric models for time $T$ and the other potential continuous auxiliary variables to make the correct specification assumption more plausible, or at least to provide a better approximation to the true model for $\pi_j(\mathbf{W}_i)$. This can be achieved by incorporating logarithmic, quadratic and higher order terms, or (finite-dimensional) B-spline terms, where the number of internal knots is fixed and does not depend on sample size $n$. Additionally, a formal goodness-of-fit procedure based on a cumulative residual process (Bakoyannis et al. 2019) can be used to provide insight about a potential violation of the model assumption for $\pi_j(\mathbf{W}, \boldsymbol{\gamma}_0)$, as it was illustrated in the HIV data analysis subsection.

By the theory of maximum likelihood estimators under misspecified models, if the model $\pi_j(\mathbf{W}_i, \boldsymbol{\gamma}_0)$ is misspecified then condition C4 still holds but with $\boldsymbol{\gamma}_0$ being replaced with $\boldsymbol{\gamma}^*$, which defines the probability that minimizes the Kullback–Leibler divergence between the true conditional distribution $\Pr(C_i = j | \Delta_i = 1, \mathbf{W}_i)$ and the assumed distribution $\pi_j(\mathbf{W}_i, \boldsymbol{\gamma}^*)$. Under this modified condition C4, the consistency in Theorem 1 holds for the parameters $\boldsymbol{\beta}_j^*$ and $\Lambda_j^*$, with $(\boldsymbol{\beta}_j^*, \Lambda_j^*) \neq (\boldsymbol{\beta}_{0,j}, \Lambda_{0,j})$, which correspond to the maximizers of the (expected) partial pseudo-likelihood under $\pi_j(\mathbf{W}, \boldsymbol{\gamma}^*)$. Similarly, $\sqrt{n}(\hat{\boldsymbol{\beta}}_{n,j} - \boldsymbol{\beta}_j^*)$, $\sqrt{n}[\hat{\Lambda}_{n,j}(t) - \Lambda_j^*(t)]$, and $\sqrt{n}[\hat{F}_{n,j}(t; \mathbf{z}_0) - F_j^*(t; \mathbf{z}_0)]$ are all asymptotically linear with influence functions given by Theorems 2–4, respectively. Consequently, the proposed estimators are still asymptotically normal, and the corresponding standard error estimators are still consistent for the true standard errors even under a misspecified model. The latter phenomenon is similar to the consistency of the sandwich variance estimator for maximum likelihood estimators under misspecified models.

The analysis of competing risks data with masked cause of failure has been considered in Craiu and Duchesne (2004). However, this method is based on a parametric cause-specific hazards model and also utilizes the computationally intensive EM-algorithm, which can be impractical for large studies such as the studies with electronic health record data. Several methods for semiparametric analysis of competing risks data with missing causes of failure have been previously proposed. Some of these methods focus on the proportional cause-specific hazards model (Goetghebeur and Ryan 1995; Lu and Tsiatis 2001; Hyun et al. 2012; Nevo et al. 2018) or the more general class of semiparametric linear transformation models (Gao and Tsiatis 2005). It has to be noted that the first stage of the analysis in the proposed approach is identical to the first stage of the multiple imputation approach for the proportional cause-specific hazards model in Lu and Tsiatis (2001). However, unlike Lu and Tsiatis (2001), we do not utilize simulation-based imputations in the second stage of the analysis and, thus, we do not introduce additional variability in the regression parameter estimates due to the finite number of imputations (Wang and Robins 1998). Therefore, as also shown empirically in the simulation studies, our regression parameter estimator is expected to be somewhat more efficient compared to the multiple imputation estimator in Lu and Tsiatis (2001). Importantly, none of the aforementioned articles provide estimators for the covariate-specific cumulative incidence functions and the corresponding standard errors. This is a significant gap in the literature, as these quantities are crucial from a clinical and implementation science perspective. Our proposed method fills this gap by proposing a unified way for inference about both the risk factor effects on the cause-specific hazards and individualized risk predictions, based on the covariate-specific cumulative incidence functions.

Among the previously proposed methods for inference about the regression coefficients under the semiparametric proportional cause-specific hazards model with missing cause of failure, the AIPW estimation method (Gao and Tsiatis 2005; Hyun et al. 2012) appears to be the most attractive approach. This is because of the so-called double robustness property that the AIPW possesses. This property ensures estimation consistency even if one of the two parametric models that are used to deal with missingness is misspecified and, also, due to their higher statistical efficiency compared to the simple inverse probability weighting estimators. However, it has been shown that if both parametric models are even slightly incorrectly specified, the AIPW estimators can yield severely biased estimates (Kang and Schafer 2007). Compared to the AIPW estimator, our proposed MPPLE estimator has the advantage of not requiring to model the probability of missingness and is also a likelihood-based approach. In the simulation studies, our proposed MPPLE was shown to be more statistically efficient compared to the AIPW estimator with a correctly specified model for the probability of missingness (in favor of the AIPW estimator). It has to be noted that this was only shown empirically in the simulation studies, and we have not formally proven this claim. In addition, the MPPLE demonstrated certain estimation robustness against misspecification of the parametric model for the failure-cause probabilities $\pi_j(\mathbf{W}, \boldsymbol{\gamma}_0)$. More importantly, inference about the infinite dimensional parameters, such as the covariate-specific cumulative incidence function, has not been studied so far in the framework of AIPW. Putting all these advantages together makes the proposed MPPLE an appealing approach to use in practice for inference under the semiparametric proportional cause-specific hazards model with missing causes of failure. A potential alternative approach would be to develop an EM-algorithm for the semiparametric proportional cause-specific hazards model. Even though this approach would be expected to be somewhat more efficient compared to our proposed MPPLE, it would be much more computationally intensive and would also be more difficult to implement in practice. The computational efficiency and ease of implementation of our MPPLE are very important characteristics in real world applications.

Although the method is illustrated with time-independent covariates, the estimator for the regression parameter presented in this paper and its properties are also valid for the case of time-dependent covariates, provided that these covariates are right-continuous with left-hand limits and of bounded variation. However, inference for the covariate-specific cumulative incidence functions with internal time-dependent covariates is trickier and requires explicit modeling of the covariate processes (Cortese and Andersen 2010). This is an interesting topic for future research. Additionally, considering nonparametric or semiparametric models for the failure-cause probabilities $\pi_j(\mathbf{W}, \boldsymbol{\gamma}_0)$ that are used to predict the missing causes of failure may be important in some applications and also interesting from a theoretical standpoint.

**Acknowledgements** We thank the Associate Editor and the two anonymous referees for their insightful comments that led to a significant improvement of this manuscript. Research reported in this publication was supported by the National Institute Of Allergy And Infectious Diseases (NIAID), Eunice Kennedy Shriver National Institute Of Child Health & Human Development (NICHD), National Institute On Drug Abuse (NIDA), National Cancer Institute (NCI), and the National Institute of Mental Health (NIMH), in accordance with the regulatory requirements of the National Institutes of Health under Award Number U01AI069911 East Africa IeDEA Consortium. The content is solely the responsibility of the authors and does not necessarily represent the official views of the National Institutes of Health. This research has also been supported by the National Institutes of Health—NIAID Grants R21 AI145662 "Estimating the cascade of HIV care under incomplete outcome ascertainment" and R01 AI102710 "Statistical Designs and Methods for Double-Sampling for HIV/AIDS" , and by the President's Emergency Plan for AIDS Relief (PEPFAR) through USAID under the terms of Cooperative Agreement No. AID-623-A-12-0001 it is made possible through joint support of the United States Agency for International Development (USAID). The contents of this journal article are the sole responsibility of AMPATH and do not necessarily reflect the views of USAID or the United States Government.